\documentclass[12pt]{article}
\usepackage{bm,amsmath,amssymb,graphicx}

\begin{document}
\pagenumbering{arabic}
\begin{titlepage}

\title{Quantum massive conformal gravity}

\author{F. F. Faria$\,^{*}$ \\
Centro de Ci\^encias da Natureza, \\
Universidade Estadual do Piau\'i, \\ 
64002-150 Teresina, PI, Brazil}

\date{}
\maketitle

\begin{abstract}
We first find the linear approximation of the second plus fourth order 
derivative massive conformal gravity action. Then we reduce the linearized 
action to separated second order derivative terms, which allows us to quantize 
the theory by using the standard first order canonical quantization method. It is 
shown that quantum massive conformal gravity is renormalizable but has ghost 
states. A possible decoupling of these ghost states at high energies is 
discussed. 
\end{abstract}

\thispagestyle{empty}
\vfill
 \par
\bigskip
\noindent * fff@uespi.br \par
\end{titlepage}
\newpage


\section{Introduction}


Massive conformal gravity \cite{Faria} is a recently developed conformal 
theory of gravity in which the gravitational action is the sum of the fourth 
order derivative Weyl action \cite{Weyl} with the second order derivative 
Einstein-Hilbert action conformally coupled to a scalar field \cite{Dirac}. 

The gravitational potential of the theory, which is composed by an attractive 
Newtonian potential and a repulsive Yukawa potential, reproduces the rotation 
curves of the major number of galaxies. In addition, the momentum space 
propagators of massive conformal gravity have a good high-energy behavior, 
which makes the theory power-counting renormalizable. However, one of these 
propagators has a negative sign between its terms, which is a common feature 
of fourth order derivative theories of gravity. In such theories this negative 
sign imply that either the energy eigenvalue spectrum is unbounded from below 
or the Hilbert space norms are negative \cite{Stelle}. Several attempts to
solve the negative energy (or negative norm) problem in higher derivative 
gravity have been carried out in the literature (see, e.g., 
\cite{Tomboulis,Salam,Antoniadis,Hawking,Narain,Biswas}).

In this paper, we analyze the consequences of the negative sign term in 
massive conformal gravity. In Sect. 2 we derive a second order derivative 
linearized massive conformal gravity action by introducing auxiliary variables. 
In Sec. 3 we canonically quantize the massive conformal gravity fields and 
show that the theory has ghosts, states with negative norm, which don't 
necessarily spoil the unitarity of the $S$-matrix. Finally, in Sec. 4 we 
present our conclusions.
    

\section{Linearized action}


Let us consider the gravitational action of massive conformal gravity, which is 
given by \footnote{This action is equivalent to the action of Ref. \cite{Faria}. 
The two actions have the same dimensions Kg.m$^{2}$/s. The only difference is that 
the action of Ref. \cite{Faria} must have the mass measured in Kg.m$^2$.}
\begin{equation}
S_{g} = \frac{1}{2kc}\int{d^{4}x} \, \sqrt{-g}\Big[\alpha \big(\varphi^{2}R 
+ 6\partial_{\mu}\varphi\partial^{\mu}\varphi\big) - \lambda^{2} 
C^{\alpha\beta\mu\nu}C_{\alpha\beta\mu\nu}  \Big],
\label{1}
\end{equation}
where $\alpha$ is a dimensionless constants, $\lambda = \hbar/mc$ 
($\hbar$ is the Planck constant and $m$ is the graviton mass), 
$k = 8\pi G/c^4$ ($G$ is the gravitational constant and $c$ is the speed of 
light in vacuum),
\begin{eqnarray}
C^{\alpha}\,\!\!_{\mu\beta\nu} &=& R^{\alpha}\,\!\!_{\mu\beta\nu} 
+ \frac{1}{2}\big( \delta^{\alpha}\,\!\!_{\nu}R_{\mu\beta} 
- \delta^{\alpha}\,\!\!_{\beta}R_{\mu\nu} + g_{\mu\beta}R^{\alpha}\,\!\!_{\nu} 
- g_{\mu\nu}R^{\alpha}\,\!\!_{\beta} \big) \nonumber \\ && 
+\frac{1}{6}\left(  \delta^{\alpha}\,\!\!_{\beta}g_{\mu\nu} 
- \delta^{\alpha}\,\!\!_{\nu}g_{\mu\beta}\right)R
\label{2}
\end{eqnarray}
is the Weyl tensor, $\varphi$ is a scalar field, 
$R^{\alpha}\,\!\!_{\mu\beta\nu}$ is the Riemann tensor, 
$R_{\mu\nu} = R^{\alpha}\,\!\!_{\mu\alpha\nu}$ is the Ricci tensor and
$R = g^{\mu\nu}R_{\mu\nu}$ is the scalar curvature. It is worth noting 
that (\ref{1}) is invariant under the conformal transformations
\begin{equation}
\tilde{g}_{\mu\nu}=e^{2\theta(x)}\,g_{\mu\nu},
\label{3} 
\end{equation}
\begin{equation}
\tilde{\varphi}=e^{-\theta(x)} \varphi,
\label{4}
\end{equation}
where $\theta(x)$ is an arbitrary function of the spacetime coordinates. 

With the help of the Lanczos identity, we can write (\ref{1}) in the form
\begin{equation}
S_{g} = \frac{1}{2kc}\int{d^{4}x} \, \sqrt{-g}\Big[\alpha
\left(\varphi^{2}R + 6 \partial_{\mu}\varphi\partial^{\mu}\varphi \right)  
- 2\lambda^{2} \left( R^{\mu\nu}R_{\mu\nu} - \frac{1}{3}R^{2}\right)\Big].
\label{5}
\end{equation}
Then, using the weak-field approximations 
\begin{equation}
g_{\mu\nu} = g^{(0)}_{\mu\nu} + h_{\mu\nu} = \eta_{\mu\nu} + h_{\mu\nu}, 
\label{6}
\end{equation}
\begin{equation}
\varphi = \varphi^{(0)}(1 + \sigma)  = \sqrt{\frac{2}{\alpha}}
(1 + \sigma),
\label{7}
\end{equation}
and keeping only the terms of second order in $h^{\mu\nu}$ and $\sigma$, we 
find that (\ref{5}) reduces to
\begin{equation}
S_{g} =  \frac{1}{kc}\int{d^{4}x} \Big[ \left( \bar{\mathcal{L}}_{EH} + 
2\sigma\bar{R} + 6 \partial_{\mu}\sigma\partial^{\mu}\sigma \right) 
- \lambda^{2}\left(\bar{R}^{\mu\nu}\bar{R}_{\mu\nu} - \frac{1}{3}\bar{R}^{2} 
\right)  \Big],
\label{8}
\end{equation}
where
\begin{equation}
\bar{R}_{\mu\nu} = \frac{1}{2} \left( \partial_{\mu}\partial^{\rho}
h_{\rho\nu} + \partial_{\nu}\partial^{\rho}h_{\rho\mu} 
- \partial_{\rho}\partial^{\rho}h_{\mu\nu} 
- \partial_{\mu}\partial_{\nu}h  \right)
\label{9}
\end{equation}
is the linearized Ricci tensor,
\begin{equation}
\bar{R} =  \partial^{\mu}\partial^{\nu}h_{\mu\nu} 
- \partial_{\mu}\partial^{\mu}h
\label{10}
\end{equation} 
is the linearized scalar curvature, and
\begin{equation}
\bar{\mathcal{L}}_{EH} = - \frac{1}{4} \Big( \partial^{\rho}
h^{\mu\nu}\partial_{\rho}h_{\mu\nu} - 2\partial^{\mu}h^{\nu\rho}
\partial_{\rho}h_{\mu\nu}+ 2\partial^{\mu}h_{\mu\nu}\partial^{\nu}h 
- \partial^{\mu}h\partial_{\mu}h  \Big)
\label{11}
\end{equation}
is the linearized Einstein-Hilbert Lagrangian density, with 
$h = \eta^{\mu\nu}h_{\mu\nu}$.

The linearized action (\ref{8}) is invariant under the coordinate gauge 
transformation
\begin{equation}
h_{\mu\nu} \rightarrow h_{\mu\nu} + \partial_{\mu}\xi_{\nu} + 
\partial_{\nu}\xi_{\mu},
\label{12}
\end{equation}
where $\xi^{\mu}$ is an arbitrary spacetime dependent vector 
field, and under the conformal gauge transformations
\begin{equation}
h_{\mu\nu} \rightarrow h_{\mu\nu} + \eta_{\mu\nu}\Lambda,
\label{13}
\end{equation}
\begin{equation}
\sigma \rightarrow \sigma - \frac{1}{2}\Lambda,
\label{14}
\end{equation}
where $\Lambda$ is an arbitrary spacetime dependent scalar field. We can fix 
these gauge freedoms by imposing the coordinate gauge condition
\begin{equation}
\partial^{\mu}h_{\mu\nu} - \frac{1}{2}\partial_{\nu}h = 0
\label{15}
\end{equation}
and the conformal gauge condition 
\begin{equation}
\bar{R}  - 6\lambda^{-2}\sigma =  0
\label{16}
\end{equation}
to (\ref{8}). However, this procedure is not suitable for the quantum 
analysis of the theory, since it introduces complication in the definition 
of the canonical commutation relations. 

Another procedure to eliminate the gauge freedoms of the theory consists on 
adding gauge fixing terms to the action such that the field equations 
obtained from the action plus the gauge fixing terms are the same as the 
gauge fixed field equations obtained from the action alone. Thus, by adding 
the gauge fixing terms \footnote{In order to simplify the calculations we 
consider the Feynman gauge in which the Lagrange multipliers entering in the 
gauge fixing terms are equal one.}
\begin{equation}
S_{GF1} = - \frac{1}{2kc}\int{d^{4}x}\left( \partial^{\mu}h_{\mu\nu} 
- \frac{1}{2}\partial_{\nu}h \right)^{2},  
\label{17}
\end{equation}
\begin{equation}
S_{GF2} =  \frac{1}{6kc}\int{d^{4}x}\left( \lambda\bar{R}  
- 6\lambda^{-1}\sigma\right)^{2},  
\label{18}
\end{equation}
to (\ref{8}), and integrating by parts, we obtain the diagonalized action
\begin{equation}
S_{d} = -\frac{1}{2kc}\int{d^{4}x} 
\Big[ \frac{1}{2}\Psi^{\mu\nu}\left( \lambda^{2}\Box - 1 \right)\Box 
\Psi_{\mu\nu} + 12\sigma\left( \Box - \lambda^{-2} \right)\sigma \Big],
\label{19}
\end{equation}
where $\Box = \partial_{\rho}\partial^{\rho}$ and
\begin{equation}
\Psi_{\mu\nu} = h_{\mu\nu} - \frac{1}{2}\eta_{\mu\nu}h.
\label{20}
\end{equation}

In order to obtain a first order canonical form, we choose the method of 
the decomposition into oscillator variables \cite{Pais} and write the 
action (\ref{19}) as
\begin{eqnarray}
S_{d} &=& \frac{1}{2kc}\int{d^{4}x} \Big[  \frac{1}{2}\Psi^{\mu\nu}
\Box \Phi_{\mu\nu} + \frac{1}{8}\lambda^{-2}\Psi_{\mu\nu}\Psi^{\mu\nu} 
- \frac{1}{4}\lambda^{-2}\Psi_{\mu\nu}\Phi^{\mu\nu} \nonumber \\ &&
+ \frac{1}{8}\lambda^{-2}\Phi_{\mu\nu}\Phi^{\mu\nu} 
- 12\sigma\left( \Box - \lambda^{-2} \right)\sigma \Big]. 
\label{21}
\end{eqnarray}
Varying this action with respect to $\Phi_{\mu\nu}$ gives
\begin{equation}
\Phi_{\mu\nu} = \Psi_{\mu\nu} - 2\lambda^{2}\Box \Psi_{\mu\nu},
\label{22}
\end{equation}
and with this the field equations obtained from action (\ref{21}) are 
equivalent to the field equations obtained from action (\ref{19}). Finally, 
with the change of variables
\begin{equation}
\Psi_{\mu\nu} = A_{\mu\nu} + B_{\mu\nu},
\label{23}
\end{equation}
\begin{equation}
\Phi_{\mu\nu} = A_{\mu\nu} - B_{\mu\nu},
\label{24}
\end{equation}
we find the action
\begin{equation}
S_{d} = \frac{1}{2kc}\int{d^{4}x} \Big[ \frac{1}{2}A^{\mu\nu}
\Box A_{\mu\nu} - \frac{1}{2}B^{\mu\nu}\left( \Box 
- \lambda^{-2} \right)B_{\mu\nu} - 12\sigma\left( \Box 
- \lambda^{-2} \right)\sigma \Big],
\label{25}
\end{equation}
which is dynamically equivalent to action (\ref{8}).

The action (\ref{25}) contains a positive energy massless spin-$2$ field 
$A_{\mu\nu}$, a negative energy massive spin-$2$ field $B_{\mu\nu}$, and a 
negative energy massive spin-$0$ field $\sigma$. Classicaly, since the theory 
is not interacting, there is no problem with the negative energy fields. 
However, when interactions are introduced instabilities can appear. Thus a 
careful analysis is necessary on the interaction of massive conformal gravity 
with matter fields, which is beyond the scope of this paper. We will deal with 
the negative energy problem at the quantum level in the next section.  


\section{Canonical quantization}


Varying the action (\ref{25}) with respect to $A^{\mu\nu}$, $B^{\mu\nu}$, and 
$\sigma$, we obtain the field equations \footnote{In this section we use 
``absolute units" in which $c=\hbar=16\pi G=1$.}
\begin{equation}
\Box A_{\mu\nu}  = 0,
\label{26}
\end{equation}
\begin{equation}
\left(\Box - m^{2} \right)B_{\mu\nu} = 0,
\label{27}
\end{equation}
\begin{equation}
\left(\Box - m^{2} \right)\sigma  = 0.
\label{28}
\end{equation}
The most general real solutions of these equations are given by 
\begin{equation}
A_{\mu\nu}(x) = \int \frac{d^{3}p}{(2\pi)^{3}} \frac{1}
{\sqrt{2\omega^{A}_{\textbf{p}}}} \sum_{r}{\left[ a^{r}_{\textbf{p}}
\epsilon^{r}_{\mu\nu}(\textbf{p})e^{ip\cdot x} 
+ c.c. \right]},
\label{29}
\end{equation}
\begin{equation}
B_{\mu\nu}(x) = \int \frac{d^{3}p}{(2\pi)^{3}}\frac{1}
{\sqrt{2\omega^{B}_{\textbf{p}}}} \sum_{s}{\left[ b^{s}_{\textbf{p}}
\varepsilon^{s}_{\mu\nu}(\textbf{p})e^{ip\cdot x} 
+ c.c. \right]},
\label{30}
\end{equation}
\begin{equation}
\sigma(x) = \int \frac{d^{3}p}{(2\pi)^{3}}\frac{1}
{\sqrt{2\omega^{\sigma}_{\textbf{p}}}} \left[ c_{\textbf{p}}
e^{ip \cdot x} + c.c. \right],
\label{31}
\end{equation}
where $\omega^{A}_{\textbf{p}} = |\textbf{p}|$, $\omega^{B}_{\textbf{p}} = 
\sqrt{|\textbf{p}|^{2} + m^{2}}$, $\omega^{\sigma}_{\textbf{p}} = 
\sqrt{|\textbf{p}|^{2} + m^{2}}$, the creation and annihilation operators 
obey the commutation relations
\begin{equation}
[a^{r}_{\textbf{p}} , a^{r'\dagger}_{\textbf{p}'}] = (2\pi)^{3}
\delta^{3}(\textbf{p} - \textbf{p}')\delta^{rr'},
\label{32}
\end{equation}
\begin{equation}
[b^{s}_{\textbf{p}} , b^{s'\dagger}_{\textbf{p}'}] = -(2\pi)^{3}
\delta^{3}(\textbf{p} - \textbf{p}')\delta^{ss'},
\label{33}
\end{equation}
\begin{equation}
[c_{\textbf{p}} , c^{\dagger}_{\textbf{p}'}] = -\frac{(2\pi)^{3}}{24}
\delta^{3}(\textbf{p} - \textbf{p}'),
\label{34}
\end{equation}
with all the other commutators equal to zero, and the polarization tensors 
satisfy the orthonormality and completeness relations
\begin{equation}
\epsilon^{r}_{\mu\nu}\epsilon^{\mu\nu r'} = \delta^{rr'}, 
\label{35}
\end{equation}
\begin{equation}
\varepsilon^{s}_{\mu\nu}\varepsilon^{\mu\nu s'} = \delta^{ss'}, 
\label{36}
\end{equation}
\begin{equation}
\sum_{r}\epsilon^{r}_{\mu\nu}\epsilon^{r}_{\alpha\beta} = \frac{1}{2} 
\left(\eta_{\mu\alpha}\eta_{\nu\beta} +\eta_{\mu\beta}\eta_{\nu\alpha} 
- \eta_{\mu\nu}\eta_{\alpha\beta} \right), 
\label{37}
\end{equation}
\begin{equation}
\sum_{s}\varepsilon^{s}_{\mu\nu}\varepsilon^{s}_{\alpha\beta} = 
\frac{1}{2} \left(\eta_{\mu\alpha}\eta_{\nu\beta} +\eta_{\mu\beta}
\eta_{\nu\alpha} - \eta_{\mu\nu}\eta_{\alpha\beta} \right). 
\label{38}
\end{equation}

We can write (\ref{25}) as
\begin{equation}
S_{d} = \int{d^{4}x} \mathcal{L},
\label{39}
\end{equation}
where
\begin{equation}
\mathcal{L} = \frac{1}{2}A^{\mu\nu}
\Box A_{\mu\nu} - \frac{1}{2}B^{\mu\nu}\left( \Box - m^{2} \right)B_{\mu\nu}  
- 12\sigma\left( \Box - m^{2} \right)\sigma
\label{40}
\end{equation}
is the massive conformal gravity Lagrangian density. Using this Lagrangian 
density, we find the canonical momenta
\begin{equation}
\Pi_{\mu\nu} = \frac{\partial\mathcal{L}}{\partial\dot{A}^{\mu\nu}} 
= -\dot{A}_{\mu\nu},
\label{41}
\end{equation}
\begin{equation}
\Theta_{\mu\nu} = \frac{\partial\mathcal{L}}{\partial\dot{B}^{\mu\nu}} 
=  \dot{B}_{\mu\nu},
\label{42}
\end{equation}
\begin{equation}
\pi = \frac{\partial\mathcal{L}}{\partial\dot{\sigma}} =  24\dot{\sigma},
\label{43}
\end{equation}
where the dot denotes the time derivative.  

It follows from (\ref{29})-(\ref{31}) and (\ref{41})-(\ref{43}) that
\begin{equation}
\Pi_{\mu\nu}(x) = -\int \frac{d^{3}p}{(2\pi)^3} \,
i\sqrt{\frac{\omega^{A}_{\textbf{p}}}{2}}\sum_{r}{\left[ 
a^{r}_{\textbf{p}}\epsilon^{r}_{\mu\nu}(\textbf{p})e^{ip\cdot x} 
+ c.c. \right]},
\label{44}
\end{equation}
\begin{equation}
\Theta_{\mu\nu}(x) =  \int \frac{d^{3}p}{(2\pi)^{3}} \,
i\sqrt{\frac{\omega^{B}_{\textbf{p}}}{2}} \sum_{s}{\left[ 
b^{s}_{\textbf{p}}\varepsilon^{s}_{\mu\nu}(\textbf{p})e^{ip\cdot x} 
+ c.c. \right]},
\label{45}
\end{equation}
\begin{equation}
\pi(x) =  24\int \frac{d^{3}p}{(2\pi)^{3}} \, i\sqrt{\frac{
\omega^{\sigma}_{\textbf{p}}}{2}} \left[ c_{\textbf{p}}e^{ip \cdot x} 
+ c.c. \right].
\label{46}
\end{equation}
Thus, by imposing the commutation rules (\ref{32})-(\ref{34}), we find
\begin{equation}
[A_{\mu\nu}(x) , \Pi_{\alpha\beta}(y)] = \frac{i}{2}
\left(\eta_{\mu\alpha}\eta_{\nu\beta} +\eta_{\mu\beta}\eta_{\nu\alpha} 
- \eta_{\mu\nu}\eta_{\alpha\beta}\right)\delta^{3}(\textbf{x} - \textbf{y}),
\label{47}
\end{equation}
\begin{equation}
[B_{\mu\nu}(x) , \Theta_{\alpha\beta}(y)] = \frac{i}{2}
\left(\eta_{\mu\alpha}\eta_{\nu\beta} +\eta_{\mu\beta}\eta_{\nu\alpha} 
- \eta_{\mu\nu}\eta_{\alpha\beta}\right)\delta^{3}(\textbf{x} - \textbf{y}),
\label{48}
\end{equation}
\begin{equation}
[\sigma(x) , \pi(y)] = i\delta^{3}(\textbf{x} - \textbf{y}),
\label{49}
\end{equation}
with all the other commutators equal to zero. 

In order to find the energy spectrum of massive conformal gravity, we need 
the Hamiltonian of the theory, which is given by
\begin{equation}
H = \int{d^{3}x} \mathcal{H},
\label{50}
\end{equation}
where
\begin{equation}
\mathcal{H} = \Pi_{\mu\nu}\dot{A}^{\mu\nu} + \Theta_{\mu\nu}\dot{B}^{\mu\nu} 
+ \pi\dot{\sigma} - \mathcal{L}
\label{51}
\end{equation}
is the massive conformal gravity Hamiltonian density. Substituting the Lagrangian 
density (\ref{40}) and the canonical momenta (\ref{41})-(\ref{43}) into 
(\ref{51}), we arrive at 
\begin{eqnarray}
\mathcal{H} &=& \frac{1}{2}\left( -\Pi_{\mu\nu}\Pi^{\mu\nu} + \partial_{i}
A_{\mu\nu}\partial^{i}A^{\mu\nu} \right) + \frac{1}{2}\big( \Theta_{\mu\nu}
\Theta^{\mu\nu} - \partial_{i}B_{\mu\nu}\partial^{i}B^{\mu\nu}
\nonumber \\ && - m^{2}B_{\mu\nu}B^{\mu\nu} \big)  + \frac{1}{2}
\Big(\frac{1}{24}\pi^{2} - 24 \partial_{i}\sigma\partial^{i}\sigma 
 - 24m^{2}\sigma^{2} \Big),
\label{52}
\end{eqnarray}
where $\partial_{i}$ denotes the spacial derivatives, with $i = 1, 2, 3$. Finally, 
after some calculation, we obtain
\begin{equation}
H = \int{\frac{d^{3}p}{(2\pi)^{3}}} \sum_{r}(\omega^{A}_{\textbf{p}} 
a^{r\dagger}_{\textbf{p}}a^{r}_{\textbf{p}}) - \int{\frac{d^{3}p}
{(2\pi)^{3}}} \sum_{s} (\omega^{B}_{\textbf{p}}b^{s\dagger}_{\textbf{p}}
b^{s}_{\textbf{p}}) - \int{\frac{d^{3}p}{(2\pi)^{3}}}
( \omega^{\sigma}_{\textbf{p}}c^{\dagger}_{\textbf{p}}c_{\textbf{p}}),
\label{53}
\end{equation}
where we have dropped the infinite constants terms that comes from 
$[a^{r}_{\textbf{p}}, a^{r\dagger}_{\textbf{p}}]$, $[b^{s}_{\textbf{p}}, 
b^{s\dagger}_{\textbf{p}}]$, and $[c_{\textbf{p}}, c^{\dagger}_{\textbf{p}}]$.

By imposing the commutation rules (\ref{32})-(\ref{34}), and using the 
Hamiltonian (\ref{53}), we find
\begin{equation}
[a^{r}_{\textbf{p}} , H] = \omega^{A}_{\textbf{p}}a^{r}_{\textbf{p}}, \ \ \ \ \ 
[a^{r \dagger}_{\textbf{p}} , H] = - \omega^{A}_{\textbf{p}}a^{r}_{\textbf{p}},
\label{54}
\end{equation}
\begin{equation}
[b^{s}_{\textbf{p}} , H] = \omega^{B}_{\textbf{p}}b^{s}_{\textbf{p}}, \ \ \ \ \ 
[b^{s \dagger}_{\textbf{p}} , H] = - \omega^{B}_{\textbf{p}}b^{s}_{\textbf{p}},
\label{55}
\end{equation}
\begin{equation}
[c_{\textbf{p}} , H] = \omega^{\sigma}_{\textbf{p}}c_{\textbf{p}}, \ \ \ \ \ 
[c^{\dagger}_{\textbf{p}} , H] = - \omega^{\sigma}_{\textbf{p}}c_{\textbf{p}}.
\label{56}
\end{equation}
If we choose the vacuum state $|0\rangle$ such that
\begin{equation}
a^{r}_{\textbf{p}} |0\rangle = 0, \ \ \ \ \ b^{s}_{\textbf{p}} |0\rangle = 0, 
\ \ \ \ \ c_{\textbf{p}} |0\rangle = 0,
\label{57}
\end{equation}
then, using (\ref{54})-(\ref{56}), we can show that the massless spin-$2$ 
state $a^{r \dagger}_{\textbf{p}}|0\rangle$ is an eigenstate of $H$ with 
energy $\omega^{A}_{\textbf{p}}$, the massive spin-$2$ state 
$b^{s \dagger}_{\textbf{p}}|0\rangle$ is an eigenstate of $H$ with energy 
$\omega^{B}_{\textbf{p}}$ and the massive spin-$0$ state 
$c^{\dagger}_{\textbf{p}}|0\rangle$ is an eigenstate of $H$ with energy 
$\omega^{\sigma}_{\textbf{p}}$. In this case, the energy spectrum of massive 
conformal gravity is bounded from below.

The Feynman propagators for the fields $A_{\mu\nu}$, $B_{\mu\nu}$ and $\sigma$ 
are, respectively,
\begin{eqnarray}
D_{A}^{\mu\nu,\alpha\beta}(x-y) &=& \langle 0 | T(A^{\mu\nu}(x)
A^{\alpha\beta}(y)) | 0 \rangle \nonumber \\ &=& -\frac{i}{2}
\left(\eta^{\mu\alpha}\eta^{\nu\beta} +\eta^{\mu\beta}\eta^{\nu\alpha} 
- \eta^{\mu\nu}\eta^{\alpha\beta}\right)
\int{\frac{d^{4}p}{(2\pi)^{4}}} \frac{e^{-ip\cdot(x-y)}}{p^2 - i\chi},
\nonumber \\ &&
\label{58}
\end{eqnarray}
\begin{eqnarray}
D_{B}^{\mu\nu,\alpha\beta}(x-y) &=& \langle 0| T(B^{\mu\nu}(x)
B^{\alpha\beta}(y)) | 0 \rangle \nonumber \\ &=&  \frac{i}{2}
\left(\eta^{\mu\alpha}\eta^{\nu\beta} +\eta^{\mu\beta}\eta^{\nu\alpha} 
- \eta^{\mu\nu}\eta^{\alpha\beta} \right)
\int{\frac{d^{4}p}{(2\pi)^{4}}} \frac{e^{-ip\cdot(x-y)}}{p^2 + m^{2} 
- i\chi}, \nonumber \\ &&
\label{59}
\end{eqnarray}
\begin{eqnarray}
D_{\sigma}(x-y) &=& \langle 0 | T(\sigma(x)
\sigma(y)) | 0 \rangle \nonumber \\ &=&  i\int{\frac{d^{4}p}
{(2\pi)^{4}}} \frac{e^{-ip\cdot(x-y)}}{p^2 + m^{2}- i\chi}, 
\label{60}
\end{eqnarray}
where $T$ denotes the time-ordered product and $\chi$ is an infinitesimal 
parameter. It follows from (\ref{23}), (\ref{58}), and (\ref{59}) that the 
propagator for the field $\Psi_{\mu\nu}$ is 
\begin{equation}
D^{\mu\nu,\alpha\beta}_{\Psi} = -\frac{i}{2}\left(\eta^{\mu\alpha}
\eta^{\nu\beta} +\eta^{\mu\beta}\eta^{\nu\alpha} - \eta^{\mu\nu}
\eta^{\alpha\beta}\right) \int{\frac{d^{4}p}
{(2\pi)^{4}}} \frac{m^{2}e^{-ip\cdot(x-x')}}{(p^2 - i\chi)
(p^2 + m^{2} - i\chi)},
\label{61}
\end{equation}
which have a good $p^{-4}$ behavior at high momenta, making massive conformal 
gravity power-counting renormalizable. 

The choice (\ref{57}) gives the required positive energy spectrum and 
renormalization but produces difficulties in the normalization of the massive 
states. Using (\ref{32})-(\ref{34}) and (\ref{57}), we find the norms
\begin{equation}
\langle 0|a^{r}_{\textbf{p}}a^{r' \dagger}_{\textbf{p}'} |0\rangle = (2\pi)^{3}
\delta^{3}(\textbf{p} - \textbf{p}')\delta^{rr'}, 
\label{62}
\end{equation}
\begin{equation}
\langle 0|b^{s}_{\textbf{p}}b^{s' \dagger}_{\textbf{p}'} |0\rangle = -(2\pi)^{3}
\delta^{3}(\textbf{p} - \textbf{p}')\delta^{ss'},
\label{63}
\end{equation}
\begin{equation}
\langle 0|c_{\textbf{p}}c^{\dagger}_{\textbf{p}'} |0\rangle = -\frac{(2\pi)^{3}}
{24}\delta^{3}(\textbf{p} - \textbf{p}'),
\label{64}
\end{equation}
and thus $b^{s \dagger}_{\textbf{p}}|0\rangle$ and 
$c^{\dagger}_{\textbf{p}}|0\rangle$ are ghost states, 
\footnote{Alternatively, if we choose $a^{r}_{\textbf{p}} |0\rangle = 0$, 
$b^{s \dagger}_{\textbf{p}} |0\rangle = 0$, and 
$c^{\dagger}_{\textbf{p}} |0\rangle = 0$, then the theory is free of ghost 
states, but the energy spectrum is unbounded from below.} which  
renders the theory non-unitary.  However, if the mass $m$ of both massive 
particles is greater than the available energy then it does not 
give rise to stable ghost states. In this case, the $S$-matrix connects 
only asymptotic states with positive norm and thus it is a unitary matrix.
Since massive conformal gravity is renormalizable, we can apply the standard
renormalization group methods to study the high-energy behavior of $m$, 
which is not a trivial thing to do. So we will leave these calculations for 
future work.


\section{Final remarks}


We have shown in this paper that massive conformal gravity is a renormalizable 
quantum theory of gravity which has two massive ghost states. A careful analysis 
is needed to check if these ghost states are decoupled from the theory 
at high energies, which would ensure the unitarity of the theory. While this 
analysis is not done we cannot rule out massive conformal gravity as a viable 
quantum theory of gravity. We are investigating this issue right now and we 
hope that this investigation help to show that quantum massive conformal gravity 
is not only renormalizable but also unitary.



\begin{thebibliography}{99}


\bibitem{Faria}
F. F. Faria, Adv. High Energy Phys. \textbf{2014}, 520259 (2014).

\bibitem{Weyl}
H. Weyl, \textit{Spacetime Matter} (Dover, New York, 1952).

\bibitem{Dirac}
P. A. M. Dirac, Proc. R. Soc. Lond. A \textbf{333}, 403 (1973).

\bibitem{Stelle}
K. S. Stelle, Phys. Rev. D \textbf{16}, 953 (1977).

\bibitem{Tomboulis}
E. Tomboulis, Phys. Lett. B \textbf{70}, 361 (1977).

\bibitem{Salam}
A. Salam and J. Strathdee, Phys. Rev. D \textbf{18}, 4480 (1978).

\bibitem{Antoniadis}
I. Antoniadis and E.T. Tomboulis, Phys. Rev. D \textbf{33}, 2765 (1986).

\bibitem{Hawking}
S.W. Hawking and T. Hertog , Phys. Rev. D \textbf{65}, 103515 (2002).  

\bibitem{Narain}
G. Narain and R. Anishetty, J. Phys. Conf. Ser. \textbf{405}, 012024 (2012).

\bibitem{Biswas}
T. Biswas, et al., Phys. Rev. Lett. \textbf{108}, 031101 (2012).

\bibitem{Pais}
A. Pais and G. E. Uhlenbeck, Phys. Rev. \textbf{79}, 145 (1950). 


\end{thebibliography}
\end{document}